\documentstyle[floats,prd,aps,epsf]{revtex}
\draft % command makes pacs numbers print

\begin{document}

%%%%%%%%%%%%%%%%%%%%%%%%%%%%%%%%%%%%%%%%%%%%%%%%%%%%%%%%%%%%%%%%%%%%%%
% uncomment the following two lines and one below for two columns!
%%%%%%%%%%%%%%%%%%%%%%%%%%%%%%%%%%%%%%%%%%%%%%%%%%%%%%%%%%%%%%%%%%%%%%

\twocolumn[\hsize\textwidth\columnwidth\hsize\csname
@twocolumnfalse\endcsname

\title{Evolving Einstein's Field Equations with Matter:\\
	The ``Hydro without Hydro'' Test}

\author{Thomas W.~Baumgarte$^{1,2}$, Scott A. Hughes$^1$ 
	and Stuart L.~Shapiro$^{1,2,3}$}

\address{${}^1$ Department of Physics, University of Illinois at
	Urbana-Champaign, Urbana, Il~61801}

\address{${}^2$ Department of Astronomy, 
	University of Illinois at Urbana-Champaign, Urbana, Il~61801}

\address{${}^3$ NCSA, 
	University of Illinois at Urbana-Champaign, Urbana, Il~61801}

\maketitle

\begin{abstract}
We include matter sources in Einstein's field equations and show that
our recently proposed 3+1 evolution scheme can stably evolve
strong-field solutions.  We insert in our code known matter solutions,
namely the Oppenheimer-Volkoff solution for a static star and the
Oppenheimer-Snyder solution for homogeneous dust sphere collapse to a
black hole, and evolve the gravitational field equations.  We find
that we can evolve stably static, strong-field stars for arbitrarily
long times and can follow dust sphere collapse accurately well past
black hole formation.  These tests are useful diagnostics for fully
self-consistent, stable hydrodynamical simulations in 3+1 general
relativity.  Moreover, they suggest a successive approximation scheme
for determining gravitational waveforms from strong-field sources
dominated by longitudinal fields, like binary neutron stars:
approximate quasi-equilibrium models can serve as sources for the
transverse field equations, which can be evolved without having to
re-solve the hydrodynamical equations (``hydro without hydro'').
\end{abstract}


\vskip2pc]

%%%%%%%%%%%%%%%%%%%%%%%%%%%%%%%%%%%%%%%%%%%%%%%%%%%%%%%%%%%%%%%%%%%%%%%
% Introduction
%%%%%%%%%%%%%%%%%%%%%%%%%%%%%%%%%%%%%%%%%%%%%%%%%%%%%%%%%%%%%%%%%%%%%%%

%\section{INTRODUCTION}

With the advent of gravitational-wave interferometry, the physics of
compact objects is entering a particularly exciting phase.  The new
generation of gravitational-wave detectors, including LIGO, VIRGO, GEO
and TAMA, may soon detect gravitational radiation directly for the
first time, opening a gravitational-wave window to the Universe and
making gravitational-wave astronomy a reality (see, {\it
e.g.},~\cite{t95}).

To learn from such observations and to dramatically increase the
likelihood of detection, one needs to predict the observed signal
theoretically.  Among the most promising sources are gravitational
waves from the coalescences of black hole and neutron star binaries.
Simulating such mergers requires self-consistent numerical solutions
to Einstein's field equations in three spatial dimensions plus time,
which is extremely challenging.  While several groups, including two
``Grand Challenge Alliances''~\cite{gc}, have launched efforts to
simulate compact binary coalescence (see
also~\cite{on97,wmm96}), the problem is far from solved.

Many of the numerical codes, including those based on the ADM
formulation of Einstein's equations (after Arnowitt, Deser and
Misner,~\cite{adm62}) develop instabilities and inevitably crash, even
for small amplitude gravitational waves on a flat background~(see,
{\it e.g.},~\cite{aetal98}).  To avoid this problem, several
hyperbolic formulations have been developed~\cite{hyperbolic}, some of
which have also been implemented numerically~\cite{sbcst97,bmsw98}.
In a recent paper~\cite{bs99} (hereafter paper I), we have modified a
formulation of Shibata and Nakamura~\cite{sn95}, and have shown that
our new formulation allows for stable, long-term evolution of
gravitational waves.  We pointed out its two advantages over many
hyperbolic formulations: it requires far fewer equations and it does
not require taking derivatives of the original 3+1 equations.  The
latter may be particularly important for the evolution of matter
sources, where matter derivatives could augment numerical error.

In this paper, we put matter sources into the field equations and test
the evolution behavior for two known strong-field solutions: the
Oppenheimer-Volkoff~\cite{ov39} solution for a static star, and the
Oppenheimer-Snyder~\cite{os39} solution for collapse of a homogeneous
dust sphere to a black hole.  We do not evolve the matter, but instead
insert the known matter solutions into the numerically evolved field
equations.  This allows us to study hydrodynamical scenarios without
re-solving hydrodynamical equations: ``hydro without hydro''.

The purpose of this Brief Report and the hydro-without-hydro approach
is twofold.  First, we demonstrate that our evolution scheme can
stably evolve strong-field solutions with matter sources.  These
calculations for strong longitudinal fields complement the tests for
wave solutions (transverse fields) presented in paper I.  This may be
an important diagnostic for overcoming stability problems in
relativistic hydrodynamical calculations ({\it cf.}~\cite{fmst99} and
discussion therein).  Second, we demonstrate that we can integrate the
field equations with {\em given} matter sources reliably.
Specifically, we show that furnishing prescribed matter sources that
already obey $\nabla\cdot{\bf T} = 0$ (where ${\bf T}$ is the matter's
stress-energy tensor) rather than self-consistently evolving the
matter together with the fields does not introduce instabilities.
This decoupling (``hydro without hydro'') suggests new possibilities
for determining gravitational waveforms emitted by, for example,
inspiraling neutron stars binaries prior to reaching the innermost
stable circular orbit by a successive approximation scheme.

In our formulation, we evolve the conformal metric
$\tilde\gamma_{ij}$, the conformal exponent $\phi$, the extrinsic
curvature's trace $K$ and conformal trace-free part $\tilde A_{ij}$,
and conformal connection functions $\tilde\Gamma^i$.  For the sake of
brevity, we refer the reader to paper I for all field equations and
their numerical implementation.

% Figure 1
\begin{figure}
\label{fig1}
\begin{center}
\leavevmode
\epsfxsize=3in
\epsffile{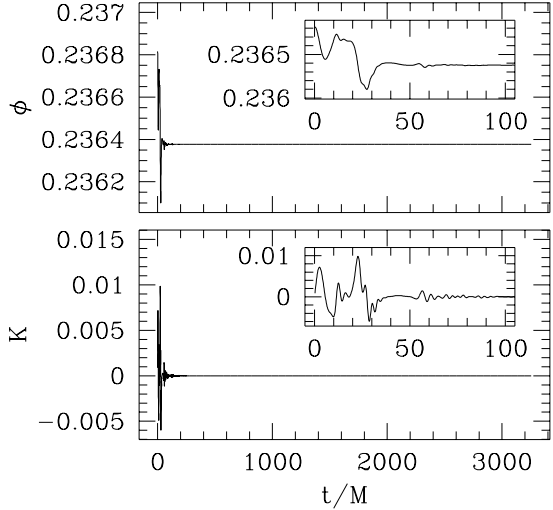}
\end{center}
\caption{Evolution of the conformal exponent $\phi$ and the trace of
the extrinsic curvature $K$ at the center for an OV static star (see
text for details).}
\end{figure}

We first study the evolution of the gravitational fields for the
Oppenheimer-Volkoff~\cite{ov39} solution of a relativistic, static
star.  We examine polytropic stellar models with equation of state $P
= \kappa \rho_0^{1 + 1/n}$, focusing on polytropic index $n = 1$.
Here $P$ is the pressure and $\rho_0$ the rest-mass density;
henceforth we set $G = 1 = c$, and we choose non-dimensional units in
which $\kappa = 1$.  We present results for a model with central
density $\rho_0^{\rm c} = 0.2$, for which the star has mass $M =
0.157$ and Schwarzschild radius $R=0.866$.  The small compaction, $R/M
= 5.5$, indicates that the star is highly relativistic.  The isotropic
radius of this star is $\bar R = 0.7$.  For comparison, the maximum
mass configuration has a central density $\rho_0^{\rm c} = 0.319$ and
a mass $M = 0.164$.

We only evolve the gravitational fields, holding the matter sources to
their OV values.  We choose zero shift ($\beta^i = 0$) and set initial
data for the lapse from the ``Schwarzschild'' lapse, $\alpha_{\rm
OV}$.  We evolve the lapse using harmonic slicing, which, for zero
shift, reduces to $\partial_t\alpha = \partial_t e^{6 \phi}$ (see,
{\it e.g.}, paper I).  We found that fixing the lapse to the exact
solution $\alpha = \alpha_{\rm OV}$ introduced instabilities, while
integrating with harmonic slicing allowed stable evolutions while
achieving the same lapse numberically.  Note that the only
non-vanishing matter sources appear in the evolution equation for $K$
[{\it cf.}\ Eq.\ (15) of paper I]: the conformal splitting explicitly
decouples transverse fields from static matter sources.

In Figure~1, we show $K$ and $\phi$ for a long-term evolution.  We
used a $(32)^3$ grid and imposed the outer boundaries at $x, y, z =
2$.  We terminated the calculation at $t = 512$ (corresponding to $t/M
\sim 3255$), and found no evidence of an instability.  Numerical noise
develops during the early part of the evolution, but this noise
propagates off the grid, and the evolution settles down into a
numerical equilibrium solution.

The numerical noise originates from {\it three} sources:
finite-difference error, noise from the surface of the star, and error
due to imposing outer boundaries at finite distance.  We now discuss
these three sources in detail.

% Figure 2
\begin{figure}
\begin{center}
\leavevmode
\epsfxsize=3in
\epsffile{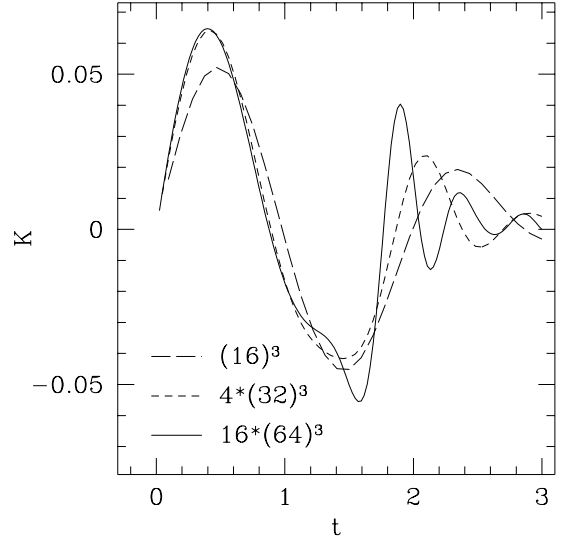}
\end{center}
\caption{Convergence test for $K$ at the center.  Note that the 
analytic solution is $K=0$.  The convergence of the scaled errors
up to $t \sim 1$ indicates second order convergence of the code.  
The breakdown of second order convergence at later times is caused
by effects of the surface of the star.}
\end{figure}

To check the local finite difference error, we perform a convergence
test and evolve the same initial data on grids of $(16)^3$, $(32)^3$
and $(64)^3$ gridpoints, all with the outer boundary at $x, y, z = 3$.
In Figure~2 we show results for $K$ at the center from these runs.
Note that the analytic solution is $K=0$.  We use second order
accurate finite difference equations, and so expect error to decrease
by a factor of four when doubling grid resolution.  This behavior is
seen at early times ($t \lesssim 1$) in Figure~2.  There are
deviations due to higher order error terms, but these decrease, and
the scaled values of $K$ converge to the second order error term.

At later times ($t \gtrsim 1$), second order convergence is spoiled by
effects from the surface of the star.  Note that the local speed of
light at the center is $dr/dt \sim \alpha/e^{2\phi}\sim 0.36 < 1$.
This delays signals from the surface, which otherwise would arrive at
$t = \bar r = 0.7$.  The code still converges, but no longer to second
order.  This effect is well known and appears in other simulations
(compare, {\it e.g.},~\cite{fmst99}).  Second-order spatial
derivatives are not smooth at the star's surface, so the
finite-difference convergence breaks down.  We have run other cases
with different stellar models (and radii) to show that this breakdown
of second order convergence is due to errors originating at the star's
surface.

% Figure 3
\begin{figure}
\begin{center}
\leavevmode
\epsfxsize=3in
\epsffile{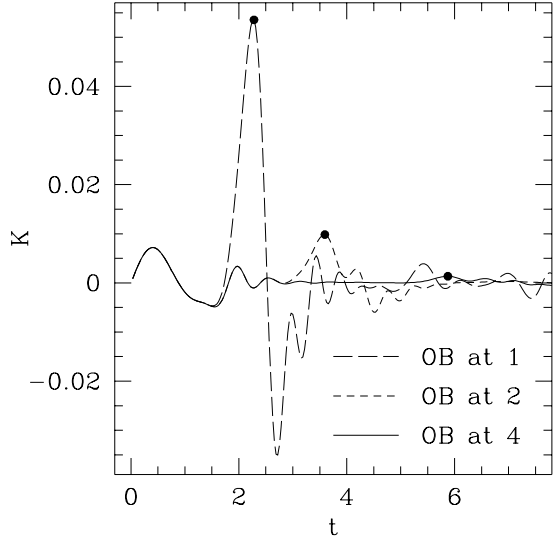}
\end{center}
\caption{Evolution of $K$ at the center for different locations
of the outer boundary.  We double the number of gridpoints
when doubling the distance to the outer boundary of the grid, so that
the grid resolution remained constant.  The dots mark the maximum
error caused by the outer boundary.}
\end{figure}

Next we analyze the effect of the outer boundary (OB).  We place the
OB at $x, y, z = 1$, $2$ and $4$, and run the code on grids of
$(16)^3$, $(32)^3$ and $(64)^3$ gridpoints, so that the resolution of
the star is constant.  The results for $K$ at the center are shown in
Figure~3.  As expected, all graphs agree until the center reaches the
domain of dependence of the different OB locations.  The run with the
OB at 1 starts to deviate from the other two runs at $t \gtrsim 1$. At
$t \gtrsim 2$ the run with OB at 2 starts to deviate from the run with
OB at 4.  The slight delay is again caused by the smaller value of the
local speed of light toward the center of the star.  The maximum error
due to the OB (marked by dots in Figure~3) quickly decreases with
larger OB location.  Since our boundary conditions take into account
the first order ($1/r$) fall-off of the fields (see paper I), one
would expect the error to scale with the square of the OB ratios, and
hence with factors of at least four in our simulations.  We find that
the errors decrease by even slightly larger factors (5.4 and 7.2).
For our resolution, the error is dominated by the local
finite-differencing error even when the outer boundary is imposed at
only a few stellar radii.

Going back to Figure~1, we can now identify the different sources of
error in the early part of the evolution.  The first peak in $K$
around $t \sim 0.4 \sim 2.5 M$ (see the panel in Figure~1) is caused
by the local finite difference error.  The next feature at $t \sim 2
\sim 12 M$, with oscillations at a higher frequency, originates from
the surface of the star.  The largest peak, at $t \sim 4 \sim 24 M$,
is caused by the outer boundary.  Reflections of these errors off the
OB reappear at later times, but ultimately propagate off the grid and
leave behind a stable numerical equilibrium solution.

%%%%%%%%%%%%%%%%%%%%%%%%%%%%%%%%%%%%%%%%%%%%%%%%%%%%%%%%%%%%%%%%%%%%%%%
% OS
%%%%%%%%%%%%%%%%%%%%%%%%%%%%%%%%%%%%%%%%%%%%%%%%%%%%%%%%%%%%%%%%%%%%%%%

%\section{Oppenheimer-Snyder Collapse}
%\label{sec4}

Turn now to an analysis of the gravitational fields associated with
the collapse of a sphere of dust, Oppenheimer-Snyder
collapse~\cite{os39}.  This configuration is highly dynamical --- the
matter very rapidly collapses to form a black hole.  This case tests
our ability to evolve into a very strong-field regime.

Again, we do not evolve the matter sources, but instead insert the
exact solution for the matter ($\rho$, $S_i$, and $S_{ij}$) into the
evolution code at each time step.  The analytic solution for
Oppenheimer-Snyder~\cite{os39} collapse is transformed into maximal
slicing and isotropic coordinates following~\cite{pst85}.  This
transformation involves only ordinary differential equations, which
can be solved numerically.  The lapse and shift corresponding to
maximal slicing and isotropic coordinates are also obtained from this
transformation and are inserted into the evolution code at each time
step.  Given the matter sources and the coordinate conditions, we
independently evolve $\phi$, $\tilde\gamma_{ij}$, $\tilde A_{ij}$, and
$\tilde\Gamma^i$ with our 3+1 code.  Having chosen maximal slicing, we
can either set $K=0$ or else evolve $K$ dynamically and check that it
indeed converges to zero.

We present results for a star that collapses from an initial
Schwarzschild radius $R_{\rm star} = 4M$ (or isotropic radius $\bar
R_{\rm star} = 2.94 M$).  The star collapses to a black hole and all
of the matter has passed inside the event horizon by $t = 12.31 M$.
We terminate the evolution at $t = 17.39 M$, the time up to which we
have constructed exact data.  The matter is so compact at the end
($\bar R_{\rm matter} \simeq 0.14 M$) that it is very poorly resolved
on our 3-D grid.  We impose the outer boundary conditions at $x = y =
z = 4M$ (in isotropic coordinates), so that initially it is quite
close to the star's surface.

% Figure 4
\begin{figure}
\begin{center}
\leavevmode
\epsfxsize=3in
\epsffile{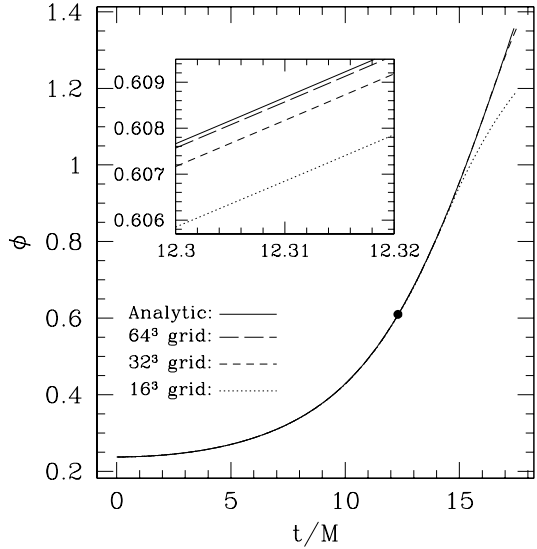}
\end{center}
\caption{ Evolution of $\phi$ at the center for different grid
resolutions for OS dust collapse to a black hole.  We hold the outer
boundary fixed at $x = y = z = 4M$ for each case.  The dot indicates
the time by which all the matter has passed inside the event horizon;
the inset is a blow-up of the curve at that time.  The numerical
evolution follows the dynamical collapse well beyond black hole
formation.  }
\end{figure}

In Figure~4, we show the evolution of the conformal exponent $\phi$ at
the origin with $(16)^3$, $(32)^3$, and $(64)^3$ gridpoints and
compare it with the exact solution.  For this run, we evolved $K$.
Figure~4 shows that the numerical evolution can follow the collapse
well past black hole formation.  The numerical solution converges to
the exact solution (see especially the inner panel of Figure~4).  At
very late times, the grid resolution becomes increasingly poor, and
ultimately convergence is spoiled by higher order finite difference
errors.  Up to these late times, the ADM and total rest mass are
reliably conserved.

As with Oppenheimer-Volkoff stars, we find that non-smoothness of the
gravitational fields at the surface spoils the second-order
convergence of quantities in the domain of dependence of the surface.
Here the effect is much stronger, since the matter density is non-zero
all the way to the surface.  However, we can lessen this effect and
improve the behavior of the evolution by imposing the maximal slicing
condition $K = 0$.  This decouples the evolution equation for
$\phi$~[{\it cf.}\ Eq.\ (14) of paper I] from the transverse fields
and from the Ricci tensor, which contains second derivatives of the
fields.  This reduces errors in the longitudinal fields that arise
from the discontinuous surface, highlighting the advantages of a
conformal splitting.

%%%%%%%%%%%%%%%%%%%%%%%%%%%%%%%%%%%%%%%%%%%%%%%%%%%%%%%%%%%%%%%%%%%%%%%
% Summary
%%%%%%%%%%%%%%%%%%%%%%%%%%%%%%%%%%%%%%%%%%%%%%%%%%%%%%%%%%%%%%%%%%%%%%%

%\section{SUMMARY AND CONCLUSIONS:\\ HYDRO WITHOUT HYDRO}
%\label{sec5}

In summary, we find that the system of equations described in paper I
accurately evolves gravitational fields associated with matter
sources.  We can evolve the fields of an Oppenheimer-Volkoff star to
extremely late times with harmonic slicing, and we can follow
Oppenheimer-Snyder collapse well beyond black hole formation into the
very strong-field regime.  We used predetermined matter sources, and
have decoupled the field evolution from hydrodynamic evolution ---
hydro without hydro.

Our findings have two important consequences.  First, the ability to
stably evolve the gravitational fields in the presence of strong-field
matter sources is an important step towards constructing fully
self-consistent, relativistic hydrodynamical codes.  Second, our tests
demonstrate that there are no fundamental difficulties evolving the
fields with prescribed matter sources obeying $\nabla\cdot{\bf T} =
0$, rather than self-consistently evolving the matter and fields
together.  This hydro-without-hydro approach suggests a possible
successive approximation scheme for calculating the gravitational-wave
signal emitted by, for example, binary neutron stars.  Outside the
innermost stable circular orbit, such binaries are dominated by
longitudinal fields and change their radial separation on a radiative
timescale which is much longer than the orbital timescale. They
therefore may be considered in quasi-equilibrium (see, {\it
e.g.},~\cite{bcsst97}).  Instead of evolving the matter
hydrodynamically, we can insert the known quasi-equilibrium binary
configuration into the field evolution code to get the transverse wave
components approximately.  Decreasing the orbital separation (and
increasing the binding energy) at the rate found for the outflow of
gravitational-wave energy would generate an approximate strong-field
wave inspiral pattern.  Such a hydro-without-hydro calculation may
yield an approximate gravitational waveform from inspiraling neutron
stars without having to couple the matter and field integrations.

%%%%%%%%%%%%%%%%%%%%%%%%%%%%%%%%%%%%%%%%%%%%%%%%%%%%%%%%%%%%%%%%%%%%%%%
% Acknowledgements
%%%%%%%%%%%%%%%%%%%%%%%%%%%%%%%%%%%%%%%%%%%%%%%%%%%%%%%%%%%%%%%%%%%%%%%

%\acknowledgments

We thank M.\ Shibata and S.\ Teukolsky for useful discussions.
Calculations were performed on SGI CRAY Origin2000 computer systems at
the National Center for Supercomputing Applications, University of
Illinois at Urbana-Champaign.  This work was supported by NSF Grant
AST 96-18524 and NASA Grant NAG 5-3420 at Illinois.

%%%%%%%%%%%%%%%%%%%%%%%%%%%%%%%%%%%%%%%%%%%%%%%%%%%%%%%%%%%%%%%%%%%%%%%
% References
%%%%%%%%%%%%%%%%%%%%%%%%%%%%%%%%%%%%%%%%%%%%%%%%%%%%%%%%%%%%%%%%%%%%%%%

\end{document}